# A Comparative Analysis of U-Net based models for Segmentation of Cardiac MRI

Ketan Suhaas Saichandran
*Department of Electrical Engineering*
*Indian Institute of Technology Roorkee*
Roorkee, India
ketansuhaas@gmail.com

*Abstract*— **Medical imaging refers to the technologies and methods utilized to view the human body and its inside, in order to diagnose, monitor, or even treat medical disorders. This paper aims to explore the application of deep learning techniques in the semantic segmentation of Cardiac short-axis MRI (Magnetic Resonance Imaging) images, aiming to enhance the diagnosis, monitoring, and treatment of medical disorders related to the heart. The focus centers on implementing various architectures that are derivatives of U-Net, to effectively isolate specific parts of the heart for comprehensive anatomical and functional analysis. Through a combination of images, graphs, and quantitative metrics, the efficacy of the models and their predictions are showcased. Additionally, this paper addresses encountered challenges and outline strategies for future improvements. This abstract provides a concise overview of the efforts in utilizing deep learning for cardiac image segmentation, emphasizing both the accomplishments and areas for further refinement.**

## I. INTRODUCTION

Magnetic Resonance Imaging (MRI) is a non-invasive imaging method that uses the principles of magnetism to produce images, which are 3-Dimensional in nature. The process includes the human body resting inside a magnetic tunnel with a steady magnetic field. This causes the hydrogen nuclei dipoles in the water present in the heart and surrounding tissues to orient themselves along the magnetic field. After a while, a sudden magnetic impulse in the form of a field is turned on due to which the dipoles disorient and gain potential energies. When the impulse is turned off the dipoles lose significant amounts of energy to produce waves, which are then used to construct 3-Dimensional images by the computer. Isolating required areas in and around the heart enables the examination of the anatomical data accurately. It also allows to eliminate extraneous scan components. Segmenting various regions of the heart is isolating the regions with different colours, as shown in *Figure 1*, to visually discriminate between them. The segmentation is performed on three classes, the right ventricle (RV), the left ventricle (LV), and the left ventricular myocardium (MLV). They require a deep examination of the anatomical and functional aspects to determine visually and quantitatively any diagnosis or anatomical defects. There are several traditional machine learning methods that were used prior to the advent of deep learning which included model-based methods or atlas-based methods. These methods required significant prior knowledge for the engineering of features that are detected in the images. As this manual construction of features and detection is not efficient enough, deep learning solves the problem by automatically constructing features, weighing them based on their priorities, and thereby identifying and detecting them in the images. Deep learning is also widely used in tasks of image classification and object detection.

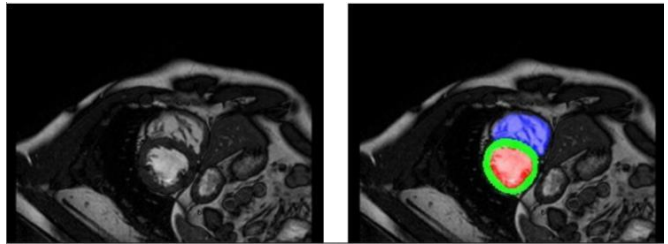

Figure 1: MRI slide (left) and predicted segmented mask (right).

## II. RELATED WORK

The literature on semantic segmentation of cardiac MRI images spans various methodologies and models designed to enhance the understanding and analysis of complex cardiac structures. The M&Ms Challenge by Campello et al. (2021) [1] introduces a comprehensive multi-centre, multi-vendor, and multi-disease cardiac segmentation initiative, providing a benchmark for advancements in the field. The U-Net architecture, pioneered by Ronneberger et al. (2015) [2], has been pivotal in biomedical image segmentation, serving as a foundational framework for subsequent innovations. In the context of cardiac MRI, Mehta and Sivaswamy (2017) [3] proposed M-Net, a Convolutional Neural Network (CNN) tailored for deep brain structure segmentation, showcasing the adaptability of segmentation techniques across diverse medical imaging domains. Attention mechanisms have proven effective in guiding segmentation networks to relevant regions, as demonstrated by Oktay et al. (2018) [4] in the Attention U-Net and by Jiang et al. (2020) [5] in the Attention M-net for automatic pixel-level micro-crack detection in photovoltaic module cells. These attention-based approaches highlight their potential applicability to the nuanced features present in cardiac MRI images. Yamanakkanavar and Lee (2021) [6] present a novel M-SegNet with a global attention CNN architecture for the automatic segmentation of brain MRI, showcasing the versatility of attention mechanisms in different imaging modalities. Isensee et al. (2018) [7] introduce nnU-Net, a self-adapting framework for U-Net-based medical image segmentation. This adaptive approach suggests promising avenues for optimizing segmentation networks based on the specific characteristics of cardiac MRI images. The optimization algorithm Adam, proposed by Kingma and Ba (2015) [8], plays a critical role in enhancing training efficiency and convergence in deep learning models, including those employed for semantic segmentation tasks in medical imaging. Model-based image segmentation methods, explored by Suetens et al. (1991) [9], provide a foundational understanding of segmentation techniques, contributing to the broader context of image analysis in medical applications. Fully Convolutional Networks (FCNs), introduced by Long et al. (2015) [10], have significantly contributed to semantic segmentation tasks, showcasing their effectiveness in various computer vision applications. Their application in cardiac MRI segmentation remains an area of ongoing exploration. This literature review emphasizes the diverse methodologies and models employed in the semantic segmentation of cardiac MRI images, showcasing the continual evolution of techniques to address the intricate structures and features within the cardiac anatomy. The references collectively highlight the progress made and underscore the potential for further advancements in this critical domain.

### III. METHODOLOGY

The dataset utilized was published by MICCAI (Medical Image Computing and Computer Assisted Interventions) for the M&Ms challenge, the University of Barcelona being the major contributor. The dataset included 368 3D MRI images in the training set out of which only 160 images are used, due to time constraints of the project, to get 1758 2D slices for our input dataset. The images had varying sizes, and hence resize operations were required to be incorporated, set at 256x256. The learning rate used was 0.001 and the batch size was 16. With ReLU activation function and dropout probability 0.5 in the hidden layers, the models were trained using Adam optimizer for 200 epochs with early stopping. Testing was performed on 25% held out data. We trained our models on a single GTX 1660Ti (6GB VRAM) laptop GPU. The U-Net was proposed specifically for the segmentation of biomedical images. It is basically an autoencoder, but with the additional skip connections which make it unique. The output of every layer of the encoder is concatenated with the inputs of the decoder blocks, preserving spatial information from the initial layers. We experimented by adding convolutional layers in the skip connections to increase feature details and named it ConvU-Net. The results were worse than the simple U-Net. The Attention U-Net, shown in *Figure 3*, was a step ahead of the traditional U-Net. The idea was to increase focus towards the regions that were more responsible for the relevant feature representations. So, to counter this, an additional feature of adding Attention Gates (as depicted in *Figure 2*) was introduced. Here, instead of concatenating the feature maps from earlier layers directly, we leverage that information to weigh the pixels of the feature maps from the upsampling layer.

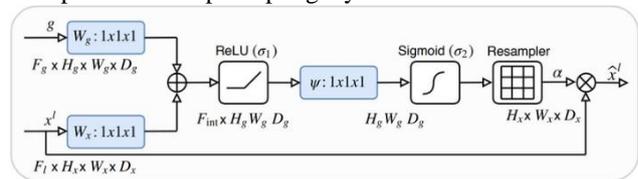

Figure 2: Attention gate (Source: Oktay et al., 2018)

These Attention Gates highlight the relevant regions in the feature maps in the skip connections before concatenating them with the decoder block inputs. The idea behind this is to calculate the attention weights for each pixel and multiply them with the pixel values. This makes the upsampled image more spatially aware.

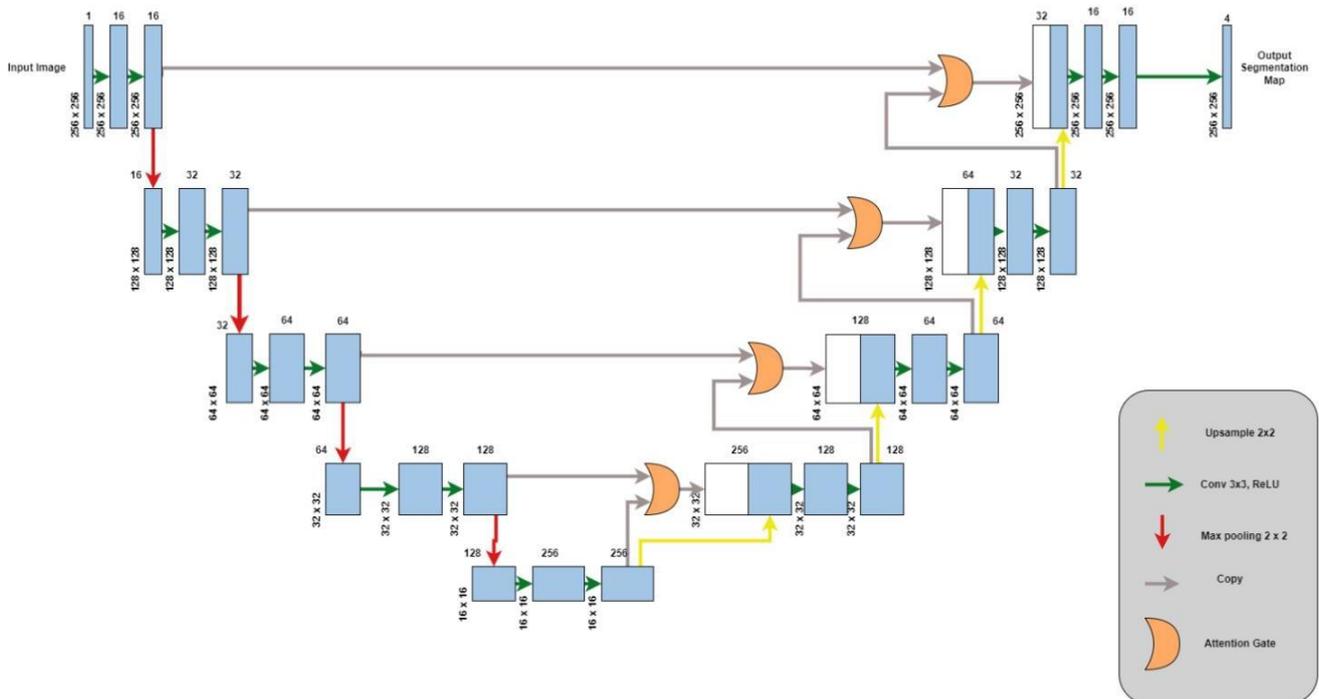

Figure 3: Attention U-Net

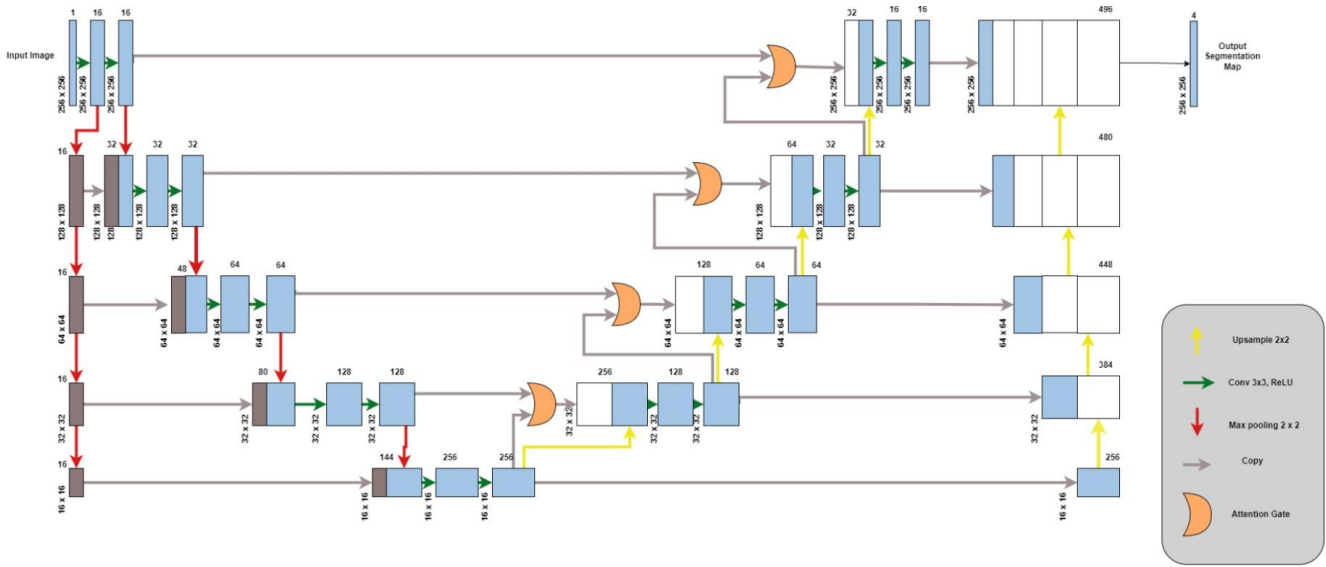

Figure 4: Attention M-Net

The differentiable nature of the Attention Gate also allows it to be trained during backpropagation, allowing the attention coefficients to get better at highlighting relevant regions.

In our implementation of the Attention M-Net (as shown in *Figure 4*), the initial feature map is down-sampled and concatenated to the inputs of every encoder block. And, the output of every decoder block is concatenated with the final layer before the final convolution.

All the models were trained on focal loss (as shown in *Figure 5*), due to its stability and adaptability. As far as the metrics are concerned, our primary focus is the Dice Coefficient (*Equation 1*).

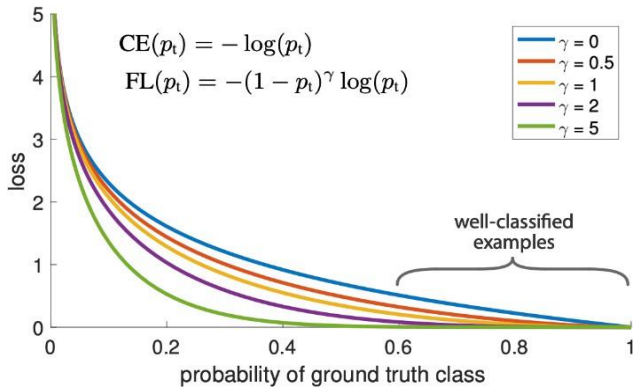

Figure 5: Focal Loss [11]

This is the Dice Coefficient (as shown in Equation 1), where p is the predicted probability of the pixels and g is the ground truth. Dice Coefficient is calculated for every class individually to assess the amount of overlap between the pixels. In the next section, the average dice coefficients of the three classes RV, LV, and MLV are presented.

$$DSC = \frac{2\sum_{i=1}^{N} p_i g_i}{\sum_{j=1}^{N} p_j^2 + g_j^2} \quad (1)$$

## IV. RESULTS AND DISCUSSIONS

The validation and test results are shown below in *Table 1*. Although the M-Net performed significantly better than the U-Net, we observed that the Attention U-Net performed better than the Attention M-Net. Additionally, adding the extra layers in M-Net consumes a high amount of memory, which could be potentially avoided.

| Method | Validation set | | Test set | |
| --- | --- | --- | --- | --- |
| | Accuracy | DSC | Accuracy | DSC |
| U-Net | 0.9943 | 0.8094 | 0.9944 | 0.755 |
| ConvU-Net | 0.9929 | 0.7517 | 0.9933 | 0.7144 |
| M-Net | 0.9947 | 0.8704 | 0.9953 | 0.8547 |
| Attention U-Net | 0.9947 | **0.876** | 0.9953 | **0.8608** |
| Attention M-Net | 0.9949 | 0.8652 | 0.9949 | 0.8394 |

Table 1: Validation and Test results

For this task, accuracy is considered to be uninformative as there is a huge overlap with the background almost all of the time, which is not implicitly considered in calculating the Dice Coefficient. The visual representation of which has been showcased using the graphs in *Figure 6* and *Figure 7*. From the graphs, we observe that the Attention U-Net has performed significantly better than the Attention M-Net.

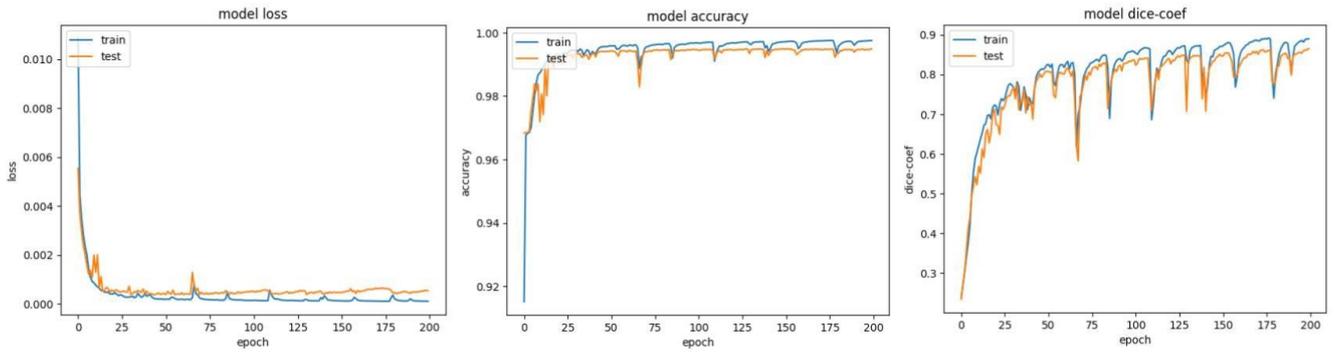

Figure 6: Loss, Accuracy, DSC vs Epochs for Attention U-Net

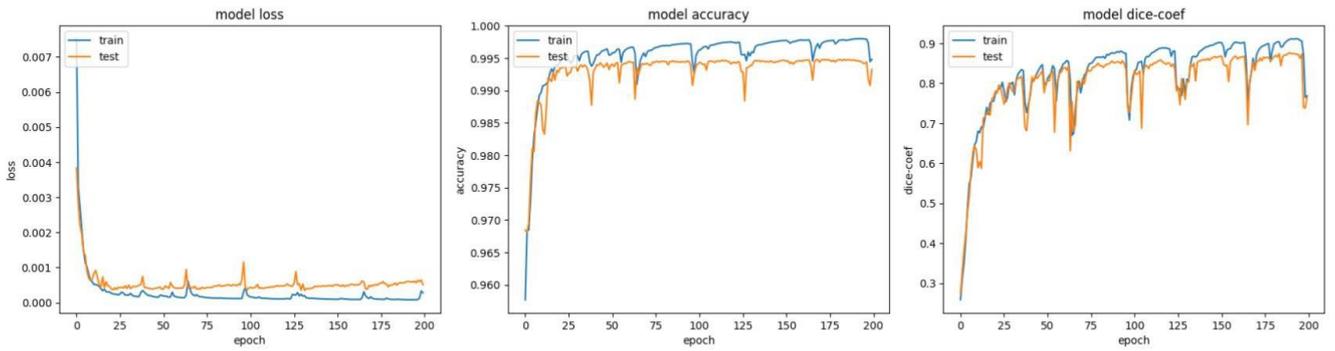

Figure 7: Loss, Accuracy, DSC vs Epochs for Attention M-Net

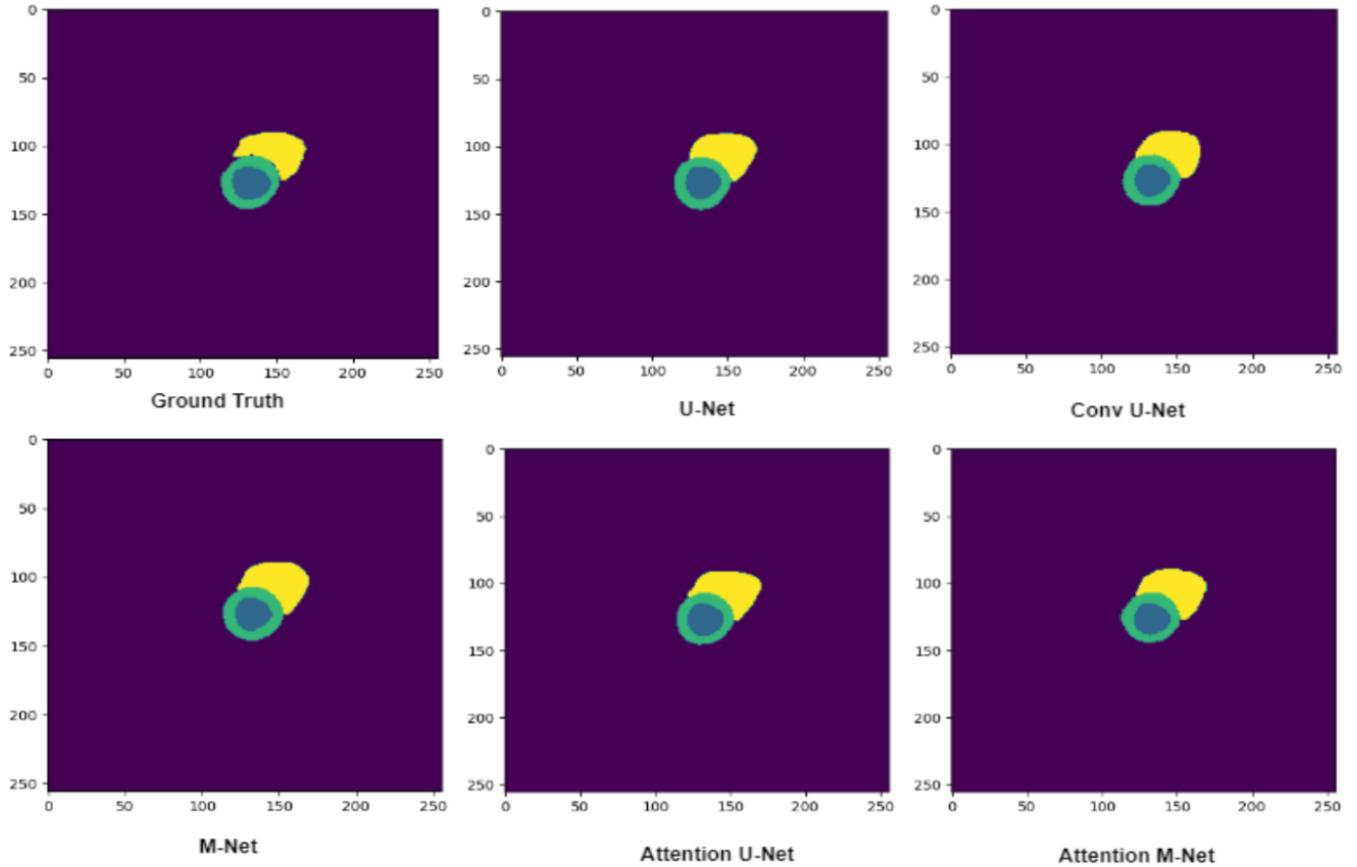

Figure 8: Segmented Masks Comparison

*Figure 8* compares the output segmented masks of all the models with the corresponding ground truth.

## V. CONCLUSION AND FUTURE DIRECTION

We have seen that the Attention U-Net is indeed a huge leap in the segmentation of cardiac MRI images from the traditional U-Net. Incorporating various attention mechanisms into image models has genuinely taken AI research to the next level, as seen in ViT (Vision Transformer) in Dosovitskiy, A. et al. (2021).

Apart from the research, it is also equally important to make it practically useful and accessible. In order to account for that, a Graphical User Interface (GUI) was created and deployed using Streamlit [12]. Features like choosing the model, uploading and displaying MRI slides, visualizing feature maps, etc were facilitated in the web application.

As part of the future direction, it will be highly valuable to explore and evaluate other attention-based models, work on larger datasets, data modalities, and more importantly create generalized models across any regional variations to be able to employ the research in a real-world setting and make it practically utilizable.

Working on this research problem can positively impact the healthcare sector significantly, paving way to treating people with diseases and disabilities in a much more efficient fashion.